\documentclass[aps,pra,amsmath,amssymb,twocolumn]{revtex4-2} 
\usepackage[dvipdfmx]{graphicx} 
\def\ch{{\cal H}}
\def\cp{{\cal P}}
\def\cq{{\cal Q}}

\def\el{{\cal L}}
\def\rv{{\vec \rho}}

\def\cl0{{\cal L}_{0}}
\def\cl1{{\cal L}_1}

\def\incl{{\hat {\cal L}}_1}
\def\hbar{\mathchar'26\mkern -9muh}

\def\v1{\mathbf{1}}

\usepackage{color}
\usepackage{hyperref} 
\hypersetup{
    bookmarks=true,         
    unicode=false,          
    pdftoolbar=true,        
    pdfmenubar=true,        
    pdffitwindow=false,     
    pdfstartview={FitH},    
    pdfkeywords={keyword1} {key2} {key3}, 
    pdfnewwindow=true,      
    colorlinks=true,       
    linkcolor=red,          
    citecolor=blue,        
    filecolor=magenta,      
    urlcolor=cyan,           
}
\begin{document}
%
%
\title{Dynamics of a quantum interacting system: \\ Extended global approach beyond the Born--Markov and secular approximations} 
\author{Chikako Uchiyama} 
\email{hchikako@yamanashi.ac.jp}
\affiliation{Faculty of Engineering, University of Yamanashi, Kofu, Yamanashi 400-8511, Japan}
\date{\today} 
\begin{abstract}
In various fields from quantum physics to biology, the open quantum dynamics of a system consisting of interacting subsystems emphasizes its fundamental functionality. The local approach, deriving a dissipator in a master equation by ignoring the inter-subsystem interaction, has been widely used to describe the reduced dynamics due to its robustness to keep the positivity of a density operator.  However, one critique is that a stationary state obtained by the approach in the limit of weak system-environment coupling is written in the form of the Gibbs state for the partial Hamiltonian by excluding the inter--subsystem interaction from the total one of the relevant system.   As an alternative, the global approach, deriving a dissipator with including the inter-subsystem interaction, under the Born--Markov and secular approximations has attracted much attention, and there is debate concerning its violation of positivity in the short--time region and/or limited parameter region for the Bohr frequencies of the subsystems. In this study, we present a formalism that leads to the time--convolutionless (time-local) master equation obtained by extending the global approach beyond the Born--Markov and secular approximations. We apply it to the excitation energy transfer between interacting sites in which only the terminal site weakly interacts with a bosonic environment of finite temperature in a manner beyond the rotating--wave approximation. We find that the formulation (1) gives the short-time behavior while preserving positivity, (2) shows the oscillatory features that the secular approximation would obscure, and (3) leads to a stationary state very near to the Gibbs state for the total Hamiltonian of the relevant system.
\end{abstract}
\maketitle
\section{Introduction}
Quantum systems that are composed of interacting subsystems are encountered in various fields including physics, chemistry, and biology\cite{Mahler,May,Amerongen}. Each of these systems inevitably interacts with its environment, and the master equation is one of the tools most widely used to extract its reduced dynamics. However, in deriving the dissipator under weak system-environment coupling, there remains controversy about the inclusion of the inter--subsystem interaction to obtain the eigenvectors of the relevant system. The distinctive approaches, including or excluding the inter--subsystem interaction, have been recently referred to as the global approach (GA) and the local approach (LA)\cite{terminology}.  

The LA, which is conventionally applied to systems in the field of quantum optics\cite{Weidlich65a,Weidlich65b,Scully67,Haken,Agarwal,Haake,Louisell,Cohen,Scully}, has been criticized as its stationary state is obtained as the Gibbs state for the partial Hamiltonian of the relevant system that excludes the inter--subsystem interaction\cite{Walls,Schwendimann,Carmichael,Gardiner,Cresser}.  In contrast, the stationary state given by the GA is the Gibbs state for the total Hamiltonian of the relevant system\cite{Walls,Schwendimann,Carmichael,Gardiner,Cresser}, which stimulates to reconceiving numerous interacting quantum systems with using the GA\cite{Carmichael74,Arimitsu,Keitel,Murao,Henrich}.  The violation of the second law of thermodynamics by the LA\cite{Novotny,Capek,Levy,Trushechkin} has also attracted much interest recently\cite{Zoubi,Scala07a,Scala08,Prosen,Rivas,Migliore,Jeske13,Santos,Joshi,Manrique,Werlang,Decordi,Hofer,Gonzalez,Seah,Raja,Naseem,Konopik,Jaschke,Camati,Benatti20,DAbbruzzo,Benatti21,Vadimov,Tupkary,Scala07b,Jeske15,Cattaneo19,Farina19,Cattaneo20,Davidovic,Trushechkin21,Farina20,Palmieri,Wichterich,Purkayastha,Purkayastha20,Scala08b,Scala09,Ribeiro,Ghoshal,Eastham,Hartmann}.

The GA generally inherits the conventional approximations for weak system--environment coupling and factorized initial conditions\cite{Redfield,Weidlich65a,Weidlich65b,Scully67,Haken,Agarwal,Haake,Louisell,Cohen,Scully,Davies,Alicki}, such as the Born--Markov approximation (BMA), with the secular approximation (SA)\cite{Novotny,Capek,Levy,Trushechkin,Zoubi,Scala07a,Scala08,Prosen,Rivas,Migliore,Jeske13,Santos,Joshi,Manrique,Werlang,Decordi,Hofer,Gonzalez,Seah,Raja,Naseem,Konopik,Jaschke,Camati,Benatti20,DAbbruzzo,Benatti21,Vadimov,Tupkary} or coarse-graining\cite{Farina20}. These approximations have conventionally been used in the microscopic derivation\cite{Breuer,Schaller} of the Gorini--Kossakowski--Sudarshan--Lindblad (GKSL) dissipator\cite{GKS,Lindblad}. 

The SA neglects the terms of the dissipator that have different eigenfrequencies, with the rationale that rapid oscillating terms in the interaction picture can be averaged out during the relaxation time of the relevant system\cite{Breuer}. This means that the SA prevents us from treating the system with similar Bohr frequencies where the eigenfrequencies are not well separated to give slower--oscillating terms which are difficult to be fully averaged out\cite{Henrich,Levy}. Moreover, because it separates the time evolution of the diagonal and off-diagonal elements of the reduced density operator in the eigenstate basis,  heat transport features that are thermodynamically inconsistent have been noted\cite{Levy,Gonzalez,Naseem,Seah,Wichterich,Purkayastha}.  Previous proposals were intended to overcome the above issues by partially\cite{Scala07b,Jeske15,Cattaneo19,Farina19,Cattaneo20,Davidovic,Trushechkin21} or fully\cite{Wichterich,Palmieri,Purkayastha,Purkayastha20} removing the SA, or by combining it with coarse-graining\cite{Farina20}, while they retained positivity with the use of the GKSL dissipator. However, another problem remains rooted in the BMA which assumes an infinitely short correlation time of the environmental variable.

With the rapid development in quantum technology, the time scale of measurements has shortened, and it now approaches the correlation time of the environmental variable\cite{Nibbering,Bigot,Banyai,Masumoto,Lodahl,Pekola}, which has opened new research avenues for environmental engineering\cite{Kwiat,Kielpinski,Viola,Kondo,BHLiu,Chiuri,ZDLiu,SYu,Khurana}. When the finiteness of the correlation time is considered, an extension beyond the BMA is desirable to describe experimentally realized situations. The extension could also provide a clue to overcoming the violation of positivity manifested beyond the SA. 

A clue is found in the dynamics described by a master equation, called the Redfield equation, which does not resort to the GKSL dissipator\cite{Munro,Suarez,Gaspard}. The Redfield equation has often been treated within the BMA that shows the violation of positivity in the short-time region for a model composed of a spin-half system interacting with a bosonic environment under a factorized initial condition\cite{Munro,Suarez,Gaspard}.   The violation of positivity could be resolved by changing the initial condition by a quantity that stems from a system--environment correlation, known as initial {\it slippage}\cite{Suarez,Gaspard,Prigogine,George}.  However, the treatment does not describe the very short--time behavior that starts from the original initial condition. A similar violation has also been reported in the case of a pair of interacting harmonic oscillators under a local interaction with a bosonic environment\cite{Farina20}.  This implies that the GA needs to be extended to include the very short--time behavior.  Then, we need to consider the finiteness of the correlation time of the environmental variables in constructing a system--environment correlation. For a finite correlation time, numerous treatments have been intensively studied to show the non-Markovian dynamics\cite{Breuer,Wolf,Breuer09,Rivas10,Chruscinski14,Rivas14,Breuer16,Clos,Guarnieri16,Alonso17,Piilo19}. 

However, the non-Markovian dynamics for the GA were mainly analyzed under the SA in \cite{Scala08b,Scala09,Ribeiro,Ghoshal}. We found recent analyses that went beyond the SA in \cite{Eastham,Hartmann} and enabled a comparison of the exact solutions of the Green function\cite{Eastham} and the exact numerical solution\cite{Hartmann}. Both analyses show a fine coincidence with the exact solution to validate the Redfield equation beyond the BMA and SA, which included the violation of positivity in a limited parameter region reported in \cite{Hartmann}. With discussions mainly limited to two non-interacting harmonic oscillators \cite{Eastham} and qubits\cite{Hartmann} immersed in a common bosonic environment, we are led to consider: how does the extension beyond the BMA and SA affect the reduced dynamics of a quantum interacting system that is partially coupled with an environment? This question is answered in the present study.

In this study, we investigate the non--Markovian dynamics of the GA beyond the SA by using a model that describes excitation energy transfer:  we consider a system composed of interacting sites, where only the terminal site interacts with a bosonic environment of finite temperature to trap excitation energy in a sink site. With the time-convolutionless (time-local) master equation (which corresponds to the Redfield equation with a time-dependent tensor), we obtain a general formula for the stationary state  as well as for the reduced dynamics up to the second-order cumulant. We find that the non--Markovian dynamics resolve the violation of positivity in the very short--time region even for the GA beyond the SA. Furthermore, we also find considerable differences in the dynamics, depending on the adoption of the SA or the BMA in the shorter--time region and/or higher environmental temperature. Although we find that the dissipator of the LA reduces to the GKSL-type, a numerical evaluation for the Ohmic spectral density shows that the trace distance between the stationary state and the expected Gibbs state is much larger than that for the GA beyond the SA.  The proposed treatment may offer a way to increase the figure of merit of a quantum device that is composed of interacting subsystems\cite{Pekola,Senior,Wiedmann,Shirai,Ptaszynski}.

This paper is organized as follows: In Section~II, we provide the formalism for treating the non--Markovian dynamics and for evaluating the stationary state. After we outline the model for the transfer of excitation energy to the sink site under local interaction with the bosonic environment in Section~III, we present our numerical results in Section~IV, a discussion in Section~V, and our conclusions in Section~VI.
\section{\label{sec:level1}Formulation}
\subsection{\label{sec:level1a}Time-convolutionless master equation}
Let us consider the reduced dynamics of the relevant system described with the time-convolutionless (time-local) master equation, which is obtained microscopically using the projection operator method\cite{Kubo,Hanggi,HSS,KTH,STH,CS,FA,US,Breuer}. We consider the Hamiltonian of the total system to be ${\ch}={\ch}_{0}+{\ch}_{1}$ with unperturbed Hamiltonian ${\ch}_{0}$ and system--environment interaction ${\ch}_{1}$, where we define the unperturbed Hamiltonian $\ch_{0}$ to be \(\ch_{0}=\ch_{\rm S}+\ch_{\rm E}\) comprising the system Hamiltonian $\ch_{\rm S}$ and the environmental Hamiltonian $\ch_{\rm E}$. Assuming a factorized initial condition between system and environment, we obtain the equation for the reduced density operator $\rho(t)$ as
\begin{equation}
 \frac{d}{dt} \rho(t) =- \frac{i}{\hbar} [\ch_{S}, \rho(t)]+ \Psi(t),
\label{eqn:21}
\end{equation}
where $\Psi(t)$ denotes the dissipator defined in time-local form as
\begin{eqnarray}
 \Psi(t) \equiv \sum_{n=2}^{\infty} \psi_{n}(t) \rho(t),
\label{eqn:22}
\end{eqnarray}
which is written with the ``ordered cumulants". A detailed derivation of Eq.~(\ref{eqn:21}) is given in Appendix~\ref{sec:levelA}.

In this study, we focus on weak-coupling of the system and environment so that we can terminate the expansion of Eq.~(\ref{eqn:21}) up to the second-order cumulant. Assuming the average of the system--environment interaction to be zero, ${\rm Tr_{E}}(\ch_{1} \rho_{{\rm E}})=0$, and introducing the equilibrium state of the environment $\rho_{{\rm E}}$, the time-convolutionless master equation with the lowest order of the dissipator becomes
\begin{eqnarray}
 \frac{d}{dt} \rho(t) =- \frac{i}{\hbar} [\ch_{S}, \rho(t)] +{\cal D}(\rho(t)) ,
\label{eqn:23} 
\end{eqnarray}
where we define $ {\cal D}(\rho(t))$ as
\begin{eqnarray}
 {\cal D}(\rho(t)) \equiv \psi_{2}(t)=\Big(\frac{i}{\hbar}\Big)^2 \int_{0}^{t} d t_{1} {\rm Tr_{E}}[ \ch_{1}, [\ch_{1}(-t_{1}), \rho_{{\rm E}}\rho(t)]],\nonumber \\
\label{eqn:24} 
\end{eqnarray}
with $\ch_{1}(t)=e^{(i/\hbar) \ch_{0} t} \ch_{1} e^{-(i/\hbar) \ch_{0} t}$.

Denoting the system--environment interaction as $\ch_{1}=\sum_{\alpha} A_{\alpha} \otimes B_{\alpha}$, we obtain
\begin{eqnarray}
{\cal D}(\rho(t)) &=& \sum_{\alpha,\beta} \int_{0}^{t} d t_{1} (\phi_{\alpha,\beta}(t_{1}) \{A_{\beta} (-t_{1}) \rho(t) A_{\alpha} \nonumber \\
&&\hspace{2.8cm} -A_{\alpha} A_{\beta} (-t_{1}) \rho(t)\}+h.c.) , \nonumber \\
\label{eqn:25} 
\end{eqnarray}
where the time evolution of $A_{\alpha}(t)$ is governed by $\ch_{\rm S}$ as $A_{\alpha}(t)=e^{\frac{i}{\hbar}\ch_{\rm S} t} A_{\alpha} e^{-\frac{i}{\hbar}\ch_{\rm S} t}$. If the relevant system includes interacting subsystems, the treatment of $A_{\alpha}(t)$ differs depending on the GA and LA. While the former uses the total system Hamiltonian $\ch_{\rm S}$ to describe the time evolution, the latter uses a part of $\ch_{\rm S}$ omitting the interaction between subsystems. In Eq.~(\ref{eqn:25}), we define the correlation function of the environment variable $B_{\alpha}$ as $\phi_{\alpha,\beta}(t)\equiv{\rm Tr_{E}}[B_{\alpha} B_{\beta} (-t) \rho_{E}]$ $\rho_{E}$, in which $B_{\alpha}(t)$ denotes the time evolution of the environment variable with $\ch_{\rm E}$ as $B_{\alpha}(t)=e^{\frac{i}{\hbar}\ch_{\rm E} t} B_{\alpha} e^{-\frac{i}{\hbar}\ch_{\rm E} t}$, and $\rho_{E}$ denotes the Gibbs state of the environment of inverse temperature $\beta(=1/k_{B} T)$ written as $\rho_{E}=e^{-\beta \ch_{\rm E}}/Z_{E}$ with partition function $Z_{E}={\rm Tr_{E}} \rho_{E}$. 

Denoting the eigenvalues and eigenstates of $\ch_{\rm S}$ as $\lambda_{n}$ and $|e_{n}\rangle$, respectively, for $n=1,\cdots,M$ and introducing the decomposition of the system operator with the completeness relation of the eigenstates as
\begin{eqnarray}
A_{\alpha}=\sum_{\epsilon} A_{\alpha}(\epsilon)=\sum_{\epsilon}
\sum_{\lambda_{m}-\lambda_{n}=\epsilon}
|e_{n}\rangle \langle e_{n}| A_{\alpha} |e_{m}\rangle \langle e_{m}|, \label{eqn:26} 
\end{eqnarray}
we obtain
\begin{eqnarray}
{\cal D}(\rho(t)) &=& \sum_{\alpha,\beta,\epsilon,\epsilon'}(\Phi_{\alpha,\beta}(\epsilon,t) \{A_{\beta}(\epsilon) \rho(t) A_{\alpha}^{\dagger}(\epsilon') \nonumber \\
&&\hspace{2.2cm}-A_{\alpha}^{\dagger}(\epsilon') A_{\beta}(\epsilon) \rho(t)\} +h.c.) , \nonumber \\
\label{eqn:27} 
\end{eqnarray}
with
\begin{eqnarray}
\Phi_{\alpha,\beta}(\epsilon,t)=\int_{0}^{t} d t_{1} \psi_{\alpha,\beta}(t_{1}) e^{\frac{i}{\hbar} \epsilon t_{1}} ,
\label{eqn:28} 
\end{eqnarray}
where we define $\psi_{\alpha,\beta}(t) \equiv{\rm Tr_{E}}[B_{\alpha}^{\dagger} B_{\beta} (-t) \rho_{E}]$.
The dissipator obtained is the precursor to the SA and the BMA, whereas the GA is often discussed under these approximations. The BMA assumes that the correlation time of the environmental variable is much shorter than the relaxation time of the system and therefore extends the upper bound of integral Eq.~(\ref{eqn:27}) to infinity, corresponding to the long-time limit. The SA frequently means taking only terms with $\epsilon=\epsilon'$ in the dissipator Eq.~(\ref{eqn:27}) under the BMA. Its physical meaning stems from the interaction picture of the master equation in which we have an extra factor $e^{\frac{i}{\hbar} (\epsilon-\epsilon') t}$ in Eq.~(\ref{eqn:27}).  The factor can be omitted as describing rapid oscillations when $(\epsilon-\epsilon')^{-1}$ is much smaller than the relaxation time of the system of interest. 

By replacing the eigenstates and eigenvalues with those associated with that part of the system Hamiltonian without the interaction terms, we obtain the dissipator of the LA, which amounts to describing only ``local" relaxations of parts of the system interacting with the environment. In Sec.~V, we discuss the difference between the GA and LA. For this purpose, we provide in the next subsection a formula for the stationary state.
\subsection{\label{sec:level1b}Stationary state}
To analyze features of the stationary state, we find that it is convenient to transform the reduced density matrix $\rho(t)$ into a Hilbert--Schmidt vector $\rv(t)=\{\rho_{11},\rho_{12},\cdots,\rho_{MM}\}$ and divide it into two parts: population (diagonal) elements, $\rv_{P}(t)$, and coherence (off-diagonal) elements, $\rv_{C}(t)$. With this decomposition, the time-local master equation, Eq.~(\ref{eqn:21}), is rewritten as
\begin{eqnarray}
\frac{d}{dt} \rv_{P}(t) &=&\Gamma_{P}(t) \rv_{P}(t) + \Gamma_{PC}(t) \rv_{C}(t), \label{eqn:vp}\\
\frac{d}{dt} \rv_{C}(t) &=&\Gamma_{CP}(t) \rv_{P}(t)+ \Gamma_{C}(t) \rv_{C}(t), \label{eqn:vc}
\end{eqnarray}
where $\Gamma_{\mu}(t)$, $\{\mu\}=\{P,PC,CP,C\}$ denote matrices with time-dependent coefficients. In the long-time limit, these matrices become time-independent as $\Gamma_{\mu} \equiv \Gamma_{\mu}(\infty)$ for $\{\mu\}=\{P,PC,CP,C\}$, corresponding to the BMA. The stationary value of the reduced density operator is a solution of the simultaneous equations subject to the time derivative being zero. Despite the simplicity of the equations, solutions are impossible when $\Gamma_{\mu}$ for $\{\mu\}=\{P,C\}$ are not invertible. There is an alternative way to find the stationary state that requires obtaining the eigenstate associated with eigenvalue zero of the matrix constructed with $\Gamma_{\mu}(\infty)$, $\{\mu\}=\{P,PC,CP,C\}$, but when the dimension is large it frequently becomes a difficult task. To overcome the difficulty, we provide a tractable formula using the final value theorem of the Laplace transform, specifically,
\begin{eqnarray}
\rv_{\mu,s}=\lim_{z \to 0} z \rv_{\mu}[z],
\label{eqn:st1} 
\end{eqnarray}
with $\rv_{\mu}[z]=\int_{0}^{\infty} \rv_{\mu}(t) e^{-z t} dt$ for $\{\mu\}=\{P,C\}$.  

Using Eq.~(\ref{eqn:vc}), we find the formal solution of coherence in the form
\begin{eqnarray}
\rv_{C}[z]=(z-\Gamma_{C})^{-1}\{\Gamma_{CP} \rv_{P}[z]+\rv_{C}(0)\}.
\label{eqn:st2} 
\end{eqnarray}
Substitution of  Eq.~(\ref{eqn:st2}) into the Laplace transform of Eq.~(\ref{eqn:vp}) yields a formal solution for the population as
\begin{eqnarray}
\rv_{P}[z]\nonumber\\
&&\hspace{-7mm}=\frac{1}{z-\Gamma_{P}
-\Gamma_{PC}\frac{1}{z-\Gamma_{C}}\Gamma_{CP}}\{\rv_{P}(0)
+\Gamma_{PC} \frac{1}{z-\Gamma_{C}}\rv_{C}(0)\},\nonumber\\
\label{eqn:st3} 
\end{eqnarray}
where $\rv_{\mu}(0)$ for $\{\mu\}=\{P,C\}$ denotes the initial condition of the population and coherence, respectively. The explicit dependence on the initial condition means we can ease the evaluation procedure if we assume that the initial condition includes only the population. We outline the numerical evaluations for our model of excitation energy transfer in Sec.~\ref{sec:level4B} for GA and Appendix~\ref{sec:levelD} for LA. We identify the coefficients for the GA and the LA using subscripts $\{\alpha\}=\{G,L\}$ respectively; i.e., $\Gamma_{\mu,\alpha}(t)$ for $\{\mu\}=\{P,PC,CP,C\}$.
\section{\label{sec:level3}Model}
We next describe the model for excitation energy transfer through the multiple interacting energy sites that has been attracting attention experimentally and theoretically\cite{Blankenship,May,Amerongen,Rebentrost,Plenio,Fischer,Trautmann,Uchiyama18,Maier}. Taking a site basis $|n\rangle$ representing a single excitation located only on the \(n\)-th component of the system, we focus on transferring excitation energy from an input energy site (\(n=1\)) to a terminal energy site (\(n=N\)). Next to the terminal site, we introduce a sink site ($n=N+1$) to trap the energy through interaction with an environment of finite temperature consisting of an infinite number of bosons. The total Hamiltonian is \(\ch=\ch_{0}+ \ch_{1}\) with \(\ch_{0}=\ch_{S}+ \ch_{E}\) where \(\ch_{S}\) represents the relevant energy-site system including the sink site and \(\ch_{E}\) the bosonic environment; explicitly, we write
\begin{eqnarray}
\ch_{S}&=& \hbar \sum_{n=1}^{N+1} \omega_{n} |n\rangle \langle n| \nonumber \\
&&\hspace{+0.2cm} + \hbar \sum_{n < m\ }^{N} V_{nm} (|n\rangle \langle m|+|m\rangle \langle n|),\label{eqn:30}\\
\ch_{E}&=& \hbar \sum_{k=1}^{\infty} \nu_{k} b_{k}^{\dagger} b_{k},
\label{eqn:31} 
\end{eqnarray}
where \(\omega_{n}\) denotes the Bohr frequency of the \(n\)-th site, \(V_{nm}\) the transition frequency between the \(n\)-th and \(m\)-th sites, \(\nu_{k}\) the frequency of the \(k\)-th boson, and \(b_{k}^{\dagger}\) (\( b_{k}\)) the creation (annihilation) boson operator of the bosonic environment. The Hamiltonian of the system--environment interaction is written as $\ch_{1}=A \otimes B$ with
\begin{eqnarray}
A&=& (|N\rangle \langle N+1|+|N+1\rangle \langle N|) , \label{eqn:32} \\
B&=& \hbar\sum_{k=1}^{\infty} g_{k} (b_{k}^{\dagger}+b_{k}), \label{eqn:33}
\end{eqnarray}
where \(g_{k}\) denotes the interaction strength between the system and the $k$-th boson of the environment. The interaction considered here goes beyond the rotating wave approximation (RWA) to include the fast counter-rotating terms in the interaction picture, the physical importance of which has been indicated previously \cite{Agarwal71,Agarwal73,Knight73,Zueco09,Fleming10,Cao11,Makela}. Let us note the difference between the RWA in Hamiltonian and the SA in the dissipator. Our formulation goes beyond both approximations. We present the numerical evaluation for this model in the next section.
\section{\label{sec:level4}Numerical Evaluation}
\subsection{\label{sec:level4A}Non-Markovian dynamics}
Using the dissipator, Eq.~(\ref{eqn:27}), we obtain the non-Markovian dynamics of the GA beyond the SA. We apply the formulation to the system with two interacting sites by setting $N=2$ in Eqs.~(\ref{eqn:30}) and (\ref{eqn:32}) where the eigenvalues of the system Hamiltonian $\ch_{\rm S}$ are $\{\lambda_{1},\lambda_{2},\lambda_{3}\} =\hbar \{\frac{1}{2}\bigl((\omega_{1}+\omega_{2})+D_{m}\bigl),\frac{1}{2} \bigl((\omega_{1}+\omega_{2}) -D_{m}\bigl),\omega_{3}\}$ with $D_{m}=\sqrt{(\omega_{1}-\omega_{2})^2+4 V_{12}^2}$, and the corresponding eigenstates $\{|e_{n}\rangle\}$ for $n=1,2,3$ are $\{[\cos \theta, \sin \theta,0]^\mathrm{T}, [-\sin \theta, \cos \theta,0]^\mathrm{T}, [0,0,1]^\mathrm{T}\}$ in the site basis; here, superscript $\mathrm{T}$ indicates the transpose operation and $\theta$ is defined to satisfy the relation $\tan 2 \theta=\frac{2 V_{12}}{\omega_{1}-\omega_{2}}$.

Focusing on the initial condition where only the first site ($N=1$) is fully excited, we find the off-diagonal elements of the density matrix between the second site ($N=2$) and the sink site ($N=3$) do not contribute to the dynamics, meaning that we can set $\rv_{P}(t)=\{\rho_{11},\rho_{22},\rho_{33}\}$ and $\rv_{C}(t)=\{\rho_{12},\rho_{21}\}$. The time-dependent coefficients for the GA, $\Gamma_{\mu,G}(t)$ for $\{\mu\}=\{P,PC,CP,C\}$, are given in Appendix~\ref{sec:levelB1}. For numerical evaluations, we set the spectral density as Ohmic, $J(\nu)=\sum_{k} g_{k}^2 \delta(\nu- \omega_{k}) \equiv s\,\nu \, e^{-\nu/\Omega_{c}}$ defining the strength of the system--environment interaction as $s$ and the cut-off frequency as $\Omega_{c}$. 

\begin{figure}[ht]
\includegraphics[scale=0.3]{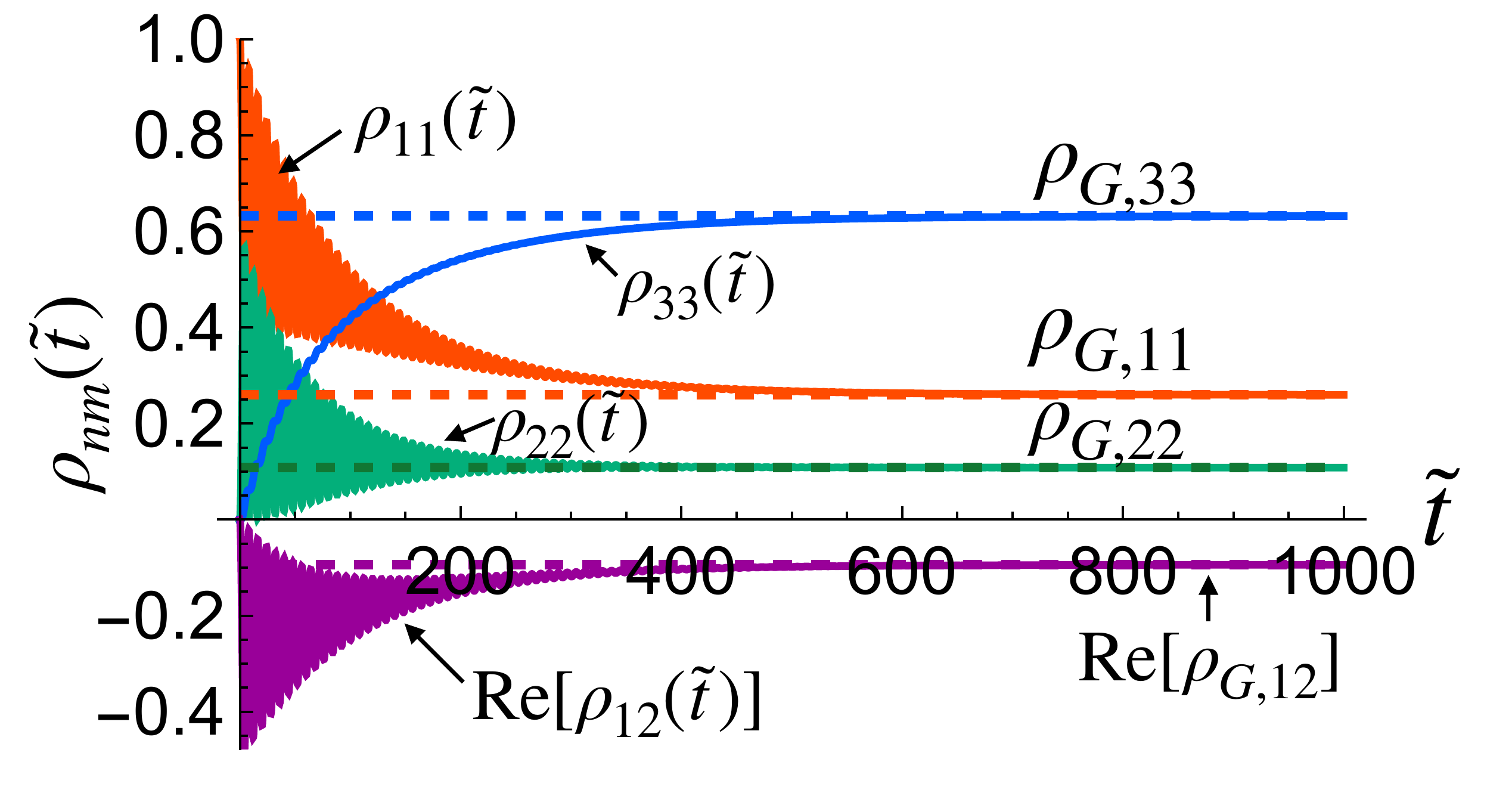}
\caption{Time evolution of each element of \(\rho({\tilde t(=\omega_{2} t} ))\) obtained by the GA beyond the SA and the BMA for the initial condition, $\rho_{11}(0)=1$, with parameter settings ${\tilde \omega}_{1}(\equiv \omega_{1}/\omega_{2})=1/2$, ${\tilde \omega}_{3}(\equiv \omega_{3}/\omega_{2})=0$, ${\tilde V}_{12}(\equiv V_{12}/\omega_{2})=3/10$, ${\tilde \Omega}_{c}(\equiv \Omega_{c}/\omega_{2})=1$, ${\tilde \beta} (\equiv \hbar \omega_{2}/(k_{B} T))=2$, and $s=1/100$. We find that the asymptotes obtained from the GA give different values, and the stationary value of coherence ${\rm Re}[\rho_{G, 12}]$ has a finite value, the features of which differ from those obtained from the LA given in Appendix~\ref{sec:levelD}.}
\label{fig:fig1}
\end{figure}

We display in Fig.~\ref{fig:fig1} the time evolution of the elements of the reduced density operator for the initial condition of the first site being fully excited, $\rho_{11}(0)=1$, where we use $\omega_{2}$ as a scaling parameter to denote the time variable ${\tilde{t}}\equiv\omega_{2} t$. Our parameter settings are given in the caption of Fig.~ \ref{fig:fig1}. These settings mean that the excitation energy at the first site is transferred through the second site with the highest energy going to the sink site with the lowest energy, approaching stationary values, $\rho_{G, nm}$ for $\{n,m\}=\{1,2,3\}$. These values are obtained using Eqs.~(\ref{eqn:st1})$\sim$(\ref{eqn:st3}), with time-independent coefficients given in Appendices~\ref{sec:levelB1} with ~\ref{sec:levelC}, and are marked as dashed lines in Fig.~\ref{fig:fig1}. We find the steady population at each site satisfying $\rho_{G,33} > \rho_{G,11} > \rho_{G,22}$; moreover, the coherence between the first and second sites, ${\rm Re}[\rho_{G, 12}]$ is finite. We discuss in the next subsection the difference between the obtained stationary states and the Gibbs states using the trace distance. 
\subsection{\label{sec:level4B}Stationary state}
\begin{figure}[ht]
\centering
\includegraphics[scale=0.2]{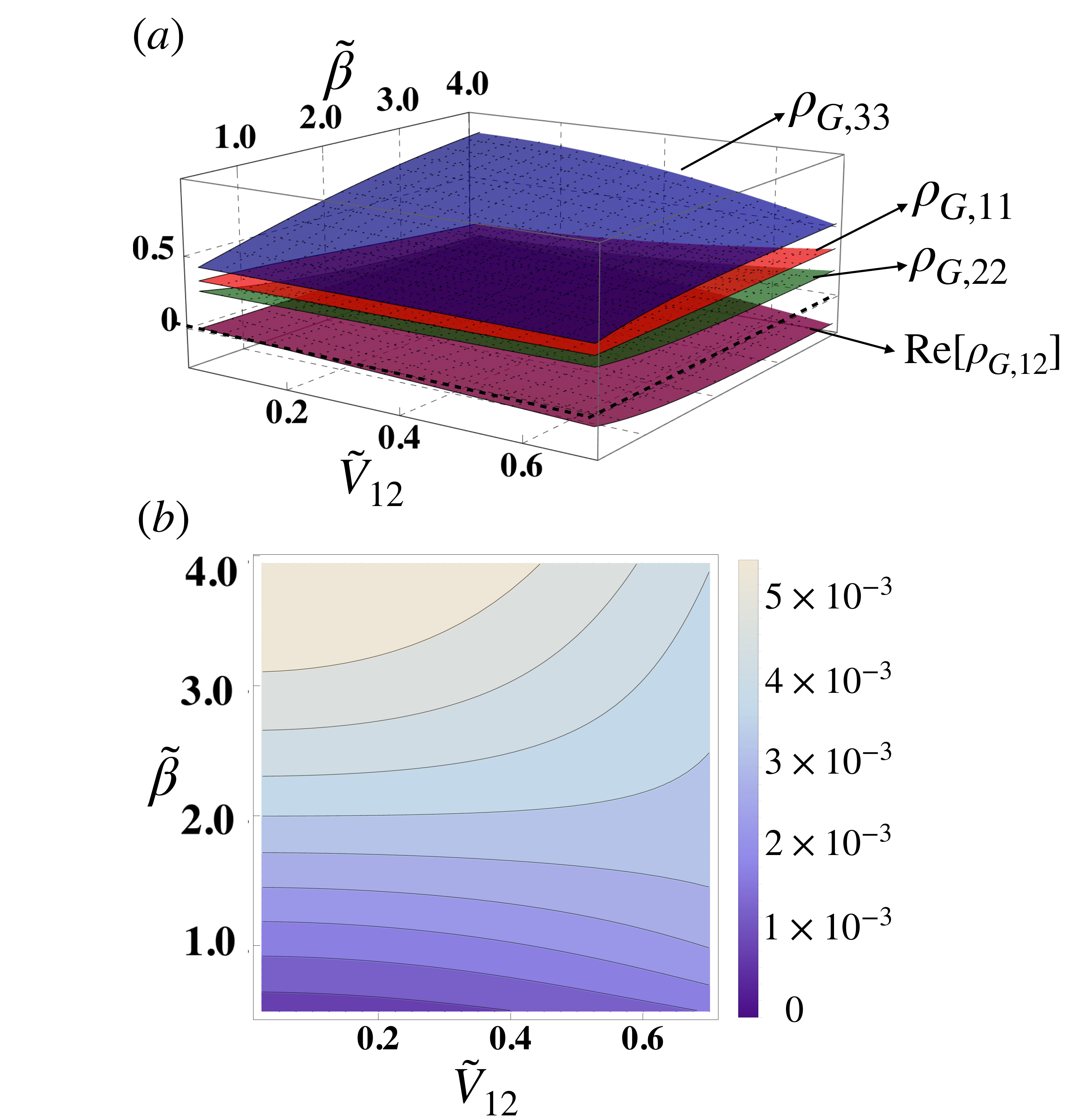}
\caption{(a) Dependence of the stationary values of the density matrix obtained by the GA beyond the SA on the inverse temperature ${\tilde \beta}$ and interaction strength between sites ${\tilde V}_{12}$. (b) Dependence of the trace distance between the stationary state $\rho_{G}$ and the expected Gibbs state $D(\rho_{G}, \rho_{{\rm Gibbs}})$. We find that this difference remains in the order of the interaction strength $O(s)$ in the range from ${\tilde \beta}$ to ${\tilde V}_{12}$ even beyond the SA. Other settings are the same as given in Fig.~\ref{fig:fig1}. }
\label{fig:fig2}
\end{figure}

We evaluate the dependence of the stationary state on inverse temperature ${\tilde \beta}$ and strength of the inter-site interaction ${\tilde V}_{12}$ for the GA [Fig.~\ref{fig:fig2}(a)] while other parameters remain unchanged from those given in Fig.~\ref{fig:fig1}. Clearly, the stationary population at the sink site, $\rho_{G,33}$, obtained using the GA, increases for lower temperatures and weaker inter-site interactions. This trend suggests that, as the inter-site interaction weakens, environmental effects become more dominant in trapping energy to the sink site, and the population $\rho_{G,33}$ increases as the temperature of the environment decreases. In contrast, as the inter-site interaction strengthens, transfers between the first and second sites become more frequent, reducing the dominance of environmental effects while elevating the absolute value of the steady coherence ${\rm Re}[\rho_{G, 12}]$.

We compare the stationary state in Fig.~\ref{fig:fig2}(a) with the Gibbs state for the total system Hamiltonian, $\rho_{{\rm Gibbs}}=e^{-\beta \ch_{\rm S}}/{\rm Z_{S}}$ with ${\rm Z_{S}}={\rm Tr_{S}}e^{-\beta \ch_{\rm S}}$, by evaluating the trace distance defined as
\begin{eqnarray}
D(\rho_{G}, \rho_{{\rm Gibbs}})=\frac{1}{2} {\rm Tr}
\bigl[\sqrt{(\rho_{G}-\rho_{{\rm Gibbs}})^2}\bigr].
\end{eqnarray}
Fig.~\ref{fig:fig2}(b) presents the dependence of $D(\rho_{G}, \rho_{{\rm Gibbs}})$ on ${\tilde \beta}$ and ${\tilde V}_{12}$ and shows that its value is of order of the strength of the system--environment interaction, $O(s)$, implying that the stationary state obtained by the GA is almost near the Gibbs state for the total Hamiltonian even beyond the SA.
\subsection{\label{sec:level4C}Effect of approximations on the GA}
Next, we show how approximations affect the dynamics obtained by the GA. We compare the non-Markovian dynamics obtained in Sec.~\ref{sec:level4} with and without the SA and/or the BMA.
\subsubsection{\label{sec:level4C1}Secular approximation}
We first study the effect of the SA on the non-Markovian dynamics by the GA, omitting the terms for $\epsilon \ne \epsilon'$ in the dissipator Eq.~(\ref{eqn:27}). The approximation excludes the oscillating terms including $e^{\pm i (\epsilon - \epsilon') t}$ in the dissipator in the interaction picture, on the assumption that the difference $\epsilon-\epsilon'$ is sufficiently large so that the terms can be averaged out during the relaxation time of the relevant system.

\begin{figure}[ht]
\includegraphics[scale=0.25]{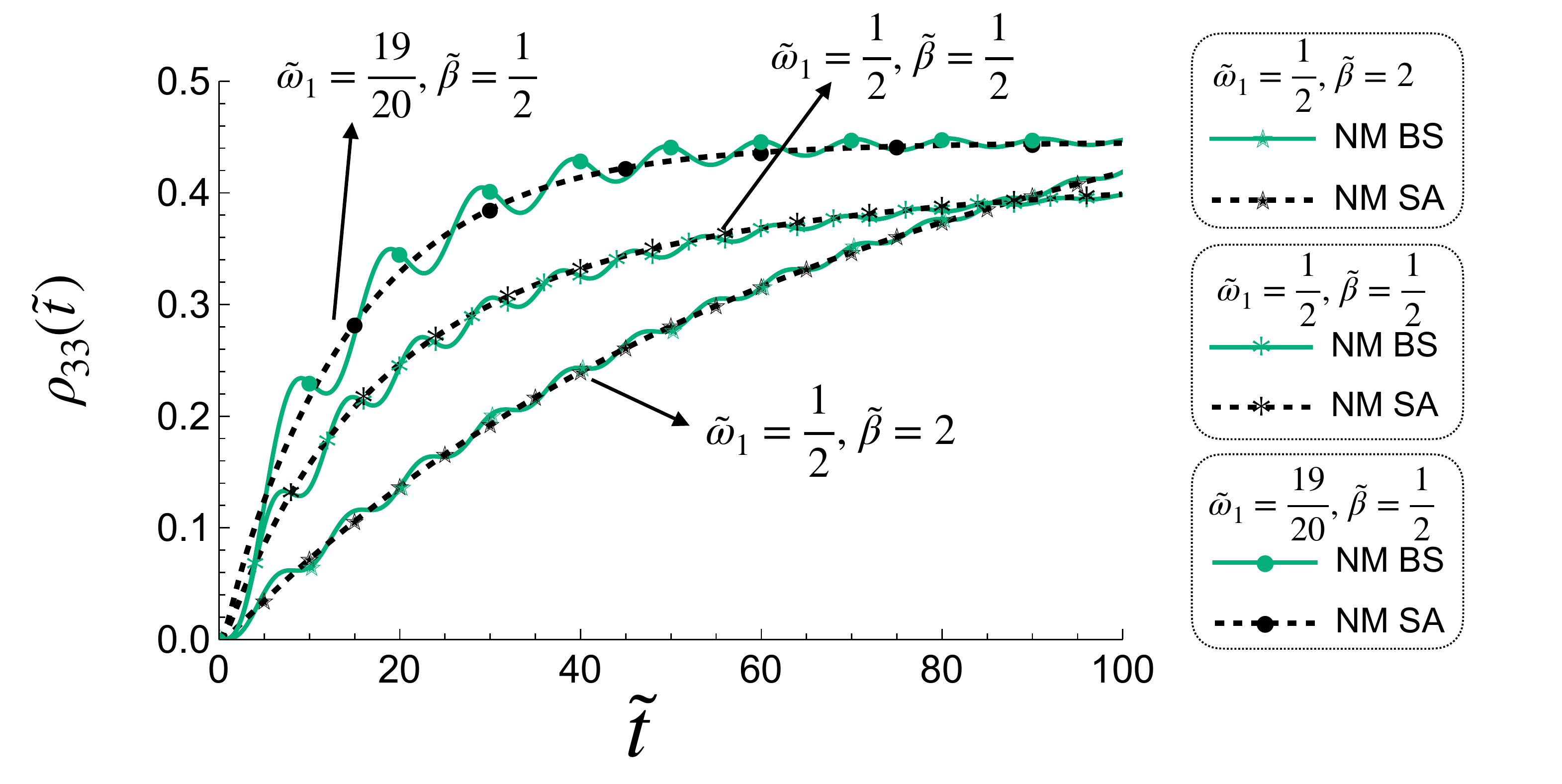}
\caption{Comparison of the non-Markovian dynamics of the sink site, $\rho_{33}({\tilde t})$, obtained by the GA with and without the SA shown as dashed and solid lines, and labeled ``NM SA" and ``NM BS", respectively. We show the dynamics for three sets of parameters $\{{\tilde \omega}_{1},{\tilde \beta}\}$ as $\{\frac{1}{2},2\}$, $\{\frac{1}{2},\frac{1}{2}\}$, and $\{\frac{19}{20},\frac{1}{2}\}$ and marked by stars, asterisks, and circles, respectively. Other settings are fixed as in Fig.~\ref{fig:fig1}. We find that the SA worsens in the short-time region and/or high temperature other than for the large difference between eigenvalues of the relevant system.}
\label{fig:fig3}
\end{figure}

Denoting the population at the sink site by $\rho_{33}({\tilde t})$, we summarize in Fig.~\ref{fig:fig3} the dynamics with and without the SA, corresponding to the dashed and solid lines labeled ``NM SA" and ``NM BS", respectively. In our model, the difference between the eigenvalues $\epsilon$ and $\epsilon'$ depends on the Bohr frequency, ${\tilde \omega}_{1}$, ${\tilde \omega}_{2}$, and the inter-system interaction ${\tilde V}_{12}$. With the contributions of these parameters being similar, we focus on trends by changing ${\tilde \omega}_{1}$ while maintaining ${\tilde V}_{12}$ and ${\tilde \omega}_{2}$ constant. By changing parameters systematically, we find that the environment temperature ${\tilde \beta}$ also contributes considerably to $\rho_{33}({\tilde t})$. To clarify the dependence on these parameters, we consider three cases by setting  $\{{\tilde \omega}_{1},{\tilde \beta}\}$ to $\{\frac{1}{2},2\}$, $\{\frac{1}{2},\frac{1}{2}\}$, and $\{\frac{19}{20},\frac{1}{2}\}$ (see Fig.~\ref{fig:fig3}). The numerical evaluations show that the SA considerably reduces the oscillation amplitude. Though these evaluations are for finite temperature, the qualitative trend of the SA is consistent with the evaluation of non-Markovianity for the model of a spin-half system interacting with a bosonic environment at zero-temperature\cite{Makela}. We find that the SA becomes more effective for low temperatures in addition to the expected region where the difference between ${\tilde \omega}_{1}$ and ${\tilde \omega}_{2}$ becomes large. However, even in this instance, we can see that, with and without the approximation, the difference is clear, particularly in the short-time region. We also reasonably find that the effect of the SA is seen for a longer time for small ${\tilde \omega}_{1}-{\tilde \omega}_{2}$ and high temperatures. This means that we need to pay attention to the effect of the SA to discuss the short-time region even if the approximation is applied to the non-Markovian dynamics. We show in the next section the effect of the BMA by comparing the trends shown in Fig.~\ref{fig:fig3}.
\subsubsection{\label{sec:level4C2}Born--Markov approximation}
The BMA describes the reduced dynamics under the effects from the environment for which the correlation time is infinitely short. However, in the microscopically derived master equation, the correlation time of the environment variables is finite which stems from the finite setting of the cut-off frequency $\Omega_{c}$ in the spectral density $J(\nu)$. We reveal the effect of the BMA on the GA for $\{{\tilde \omega}_{1},{\tilde \beta}\}=\{\frac{1}{2},2\}$ by comparing the non-Markovian dynamics shown in Fig.~\ref{fig:fig3}.

\begin{figure}[ht]
\includegraphics[scale=0.27]{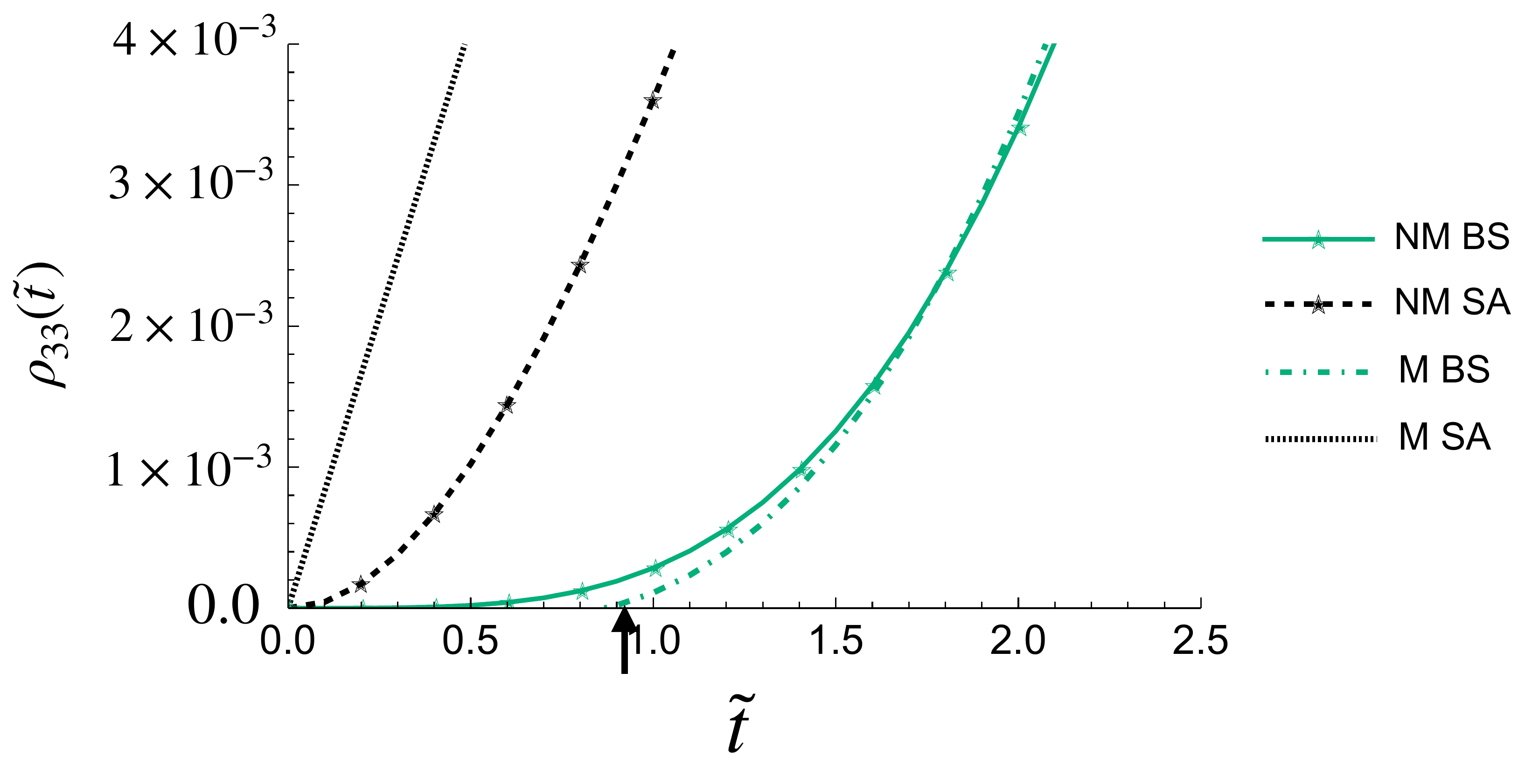}
\caption{Comparison of the dynamics of the sink site, $\rho_{33}({\tilde t})$, obtained by the GA with and without the BMA and/or the SA for $\{{\tilde \omega}_{1},{\tilde \beta}\}=\{\frac{1}{2},2\}$; the other parameters are fixed as in Fig.~\ref{fig:fig1}. The dotted and dot-dashed lines, labeled ``M SA" and ``M BS", respectively, show the dynamics under the BMA with and without the SA. The arrow below the horizontal axis indicates the point where the sign of the values from ``M BS" changes from negative to positive. For comparison, we show the non-Markovian dynamics with and without the SA (dashed and solid lines labeled ``NM SA" and ``NM BS" respectively.}
\label{fig:fig4}
\end{figure}

We present the dynamics of $\rho_{33}({\tilde t})$ under the application of both the BMA and the SA (Fig.~\ref{fig:fig4}; dotted line labeled ``M SA"), which is frequently taken as the GA. Combining the approximations reduces the dissipator to the GKSL form. We find that the dynamics show their fastest increase without the ``inertial" feature compared with the non-Markovian dynamics. Moreover, when we remove the SA, the dynamics for which are marked by the dot-dashed line labeled ``M BS" in Fig.~\ref{fig:fig4}, we find violations of positivity in a very-short-time region up to the time indicated by an arrow. The violation is not surprising because the finiteness of the correlation time does not match the condition for applying the BMA, which is consistent with conventional studies\cite{Suarez,Gaspard}. For comparison, we show the non-Markovian dynamics with and without the SA (dashed and solid lines labeled ``NM SA" and ``NM BS") shown in Fig.~\ref{fig:fig3}. We find that the violation of positivity is resolved by including the non-Markovian effect.

\begin{figure}[ht]
\includegraphics[scale=0.165]{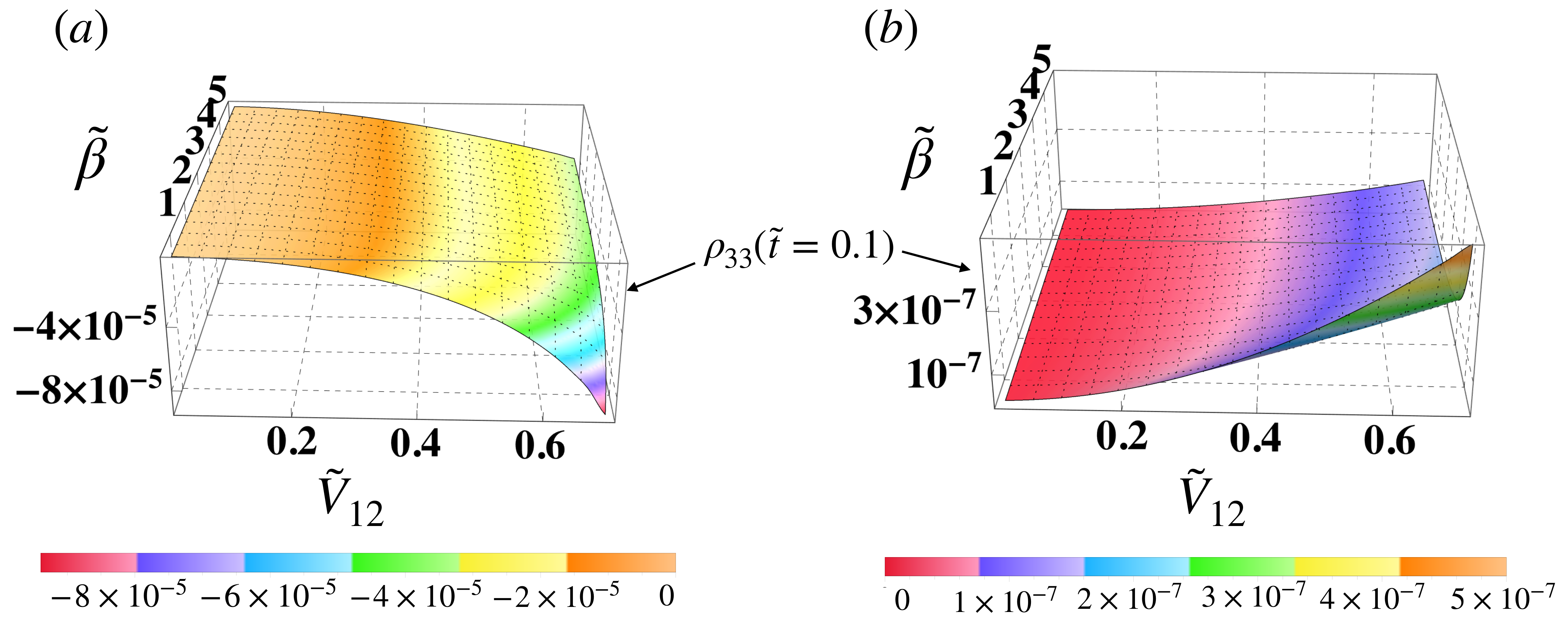}
\caption{Dependence of $\rho_{33}({\tilde t}=0.1)$ obtained by the GA beyond the SA on the inverse temperature ${\tilde \beta}$ and the interaction strength between sites ${\tilde V}_{12}$; all other parameters are fixed as in Fig.~\ref{fig:fig1}. (a) under BMA beyond the SA (``M BS") the values are negative, and (b) under non-Markovian dynamics beyond the SA (``NM BS") the values are strictly positive. We find that the positivity in the short-time region is recovered by the treatment ``NM BS".}
\label{fig:fig5}
\end{figure}

To discuss whether the violation of positivity in $\rho_{33}({\tilde t})$ under the BMA beyond the SA (``M BS") is limited to the parameter settings specified in Fig.~\ref{fig:fig4}, we display in Fig.~\ref{fig:fig5}(a) the dependence of $\rho_{33}({\tilde t}=0.1)$ on the inverse temperature ${\tilde \beta}$ and the interaction strength between sites ${\tilde V}_{12}$. We find that the values are negative throughout the parameter region, which is the opposite to that obtained by the non-Markovian dynamics beyond the SA [Fig.~\ref{fig:fig5}(b), curve ``NM BS"].
From the evaluations shown in Figs.~\ref{fig:fig1}--\ref{fig:fig5}, we find that the non-Markovian dynamics for the GA beyond the SA show the stationary state of the Gibbs state for the total system Hamiltonian without positivity violation in the very short time region. However, because non-Markovian dynamics reduce to the dynamics of the BMA in the long-time region, this approximation in the low temperature region is an issue needing further study \cite{Carmichael99, Rivas}, which we discuss in the next section.

\section{\label{sec:level8}Discussion}
Using the time-convolutionless master equation, this study demonstrates that the extension of the GA beyond the BMA and the SA affects the reduced dynamics of a quantum interacting system.  It preserves positivity in the short time region and describes the stationary state as close to the Gibbs state of the total Hamiltonian of the relevant system. 
Several topics remain for further discussion, as follows.

First, we discuss the difference between the GA and the LA. Because the latter assumes that the inter--subsystem interaction does not contribute to the dissipation, the dissipator, Eq.~(\ref{eqn:24}), is obtained by excluding the inter--site interaction $V_{12}$ from Eq.~(\ref{eqn:30}), which provides the time-dependent coefficients as expressed in Appendix~\ref{sec:levelB2}. Considering the Born--Markovian (long-time) limit, we obtain an analytic form of the stationary state for the LA beyond the SA, Eq.~(\ref{eqn:D1}), which is different from the Gibbs state for the total Hamiltonian of the relevant system (see Fig.~\ref{fig:figD2}).

The next issue is the positivity of time evolution obtained by the GA beyond the SA. In the analysis presented in Section~IV, we showed that the violation of positivity in the short-time region is resolved by including the non--Markovian effect. However, for the stationary state, we find the parameter region in which the smallest eigenvalue of the stationary state becomes negative, corresponding to low temperatures and strong inter-site interaction for large Bohr frequency of site 1  (see Appendix~\ref{sec:levelE}). We find two possible sources of negativity: in assuming an Ohmic spectral density\cite{Carmichael99, Rivas} and in the accuracy of the master equation when taken to the second-order cumulant\cite{Fleming,Mori}. 

Concerning the assumption on the spectral density, we find a recurrent theme in the lengthening of the correlation time of the bosonic environment for low temperatures for the Ohmic spectral density, which does not match the assumptions of the BMA\cite{Carmichael99, Rivas}. Becuase a negative stationary value is indicated for the long correlation time of the phenomenological exponential correlation function for the GA beyond the SA\cite{Hartmann}, and because the non--Markovian dynamics eventually reduce to the dynamics under BMA in the long-time region, we regard it as a source of negativity. 

For accuracy in the stationary values of the second-order master equation, unbalanced accuracy between the population and coherence is indicated\cite{Fleming,Mori}. To overcome this difficulty, an analytic continuation for approaching the second order of the mean-force Gibbs state is proposed\cite{Thingna12,Thingna13}. We find its recent extension to the master equation adopting the SA\cite{Thingna22}, but the negative stationary values remain. This suggests that higher-order cumulants are necessary, as was concluded in \cite{Laird,Breuer,US}.

As the GA beyond the SA gives the non-GKSL dissipator, Eq.~(\ref{eqn:27}), there may be concerns with its positivity violation. For the reduced dynamics of two non-interacting bosonic modes\cite{Eastham} or qubits\cite{Hartmann} that are immersed in a common bath, the numerical evaluation of Eq.~(\ref{eqn:27}) matches the analytic or numerical solution. This suggests that the non-GKSL dissipator itself is not the only reason for the violation of positivity.  Indeed, recent studies on canonical forms prove that Eq.~(\ref{eqn:27})  could be transformed into a GKSL dissipator with time-dependent coefficients\cite{Hall14}.  However, because the transformation does not guarantee positivity and even complete positivity, the authors of \cite{Hall14} also mention the need to identify the region satisfying the legitimate dynamics as shown in \cite{Hall14,Hall08,Maniscalco07} for the spin-half case.  This case corresponds to the numerical evaluation in Appendix~\ref{sec:levelE}, although we discuss only the positivity aspect.

Finally, we discuss the possibility of simulating the model presented in this paper for validation. The candidates are roughly divided into $\langle 1\rangle$ the analytically or $\langle 2\rangle$ numerically exact methods. For $\langle 1\rangle$, the two following possibilities exist: (1) the Heisenberg equations of motion and (2) the pseudomode model.  The application of (1) produces a finite set of coupled differential equations for two non-interacting harmonic oscillators that are immersed in a common bosonic environment within the RWA in \cite{Eastham}.  We find the set for our model to be infinite, and this precludes us from obtaining an exact solution. The application of (2) to the two non-interacting qubits that commonly couple with a bosonic environment beyond the RWA enabled the authors of \cite{Hartmann} to replicate the dynamics obtained from the Redfield (time-convolutionless) equation.  Their replication was based on the Lorentzian-like spectral density to describe system--environment interaction, according to \cite{Garraway,Imamoglu}. However, we find that we need an infinite number of the pseudomodes to replicate the dynamics of our model based on the Ohmic spectral density\cite{Tamascelli}, and this precludes us from applying the model. 

For $\langle 2\rangle$, some typical examples are (1) the multilayer formulation of the multiconfiguration time-dependent Hartree theory (ML-MCTDH)\cite{Wang01,Wang03}, (2) the quasi adiabatic path integral method (QUAPI)\cite{Makri}, (3) the hierarchical equations of motion (HEOM)\cite{Tanimura89,Tanimura90,Nakamura} and its variants(eHEOM)\cite{Tang}, and (4) the hierarchy of pure states (HOPS)\cite{Hartmann20,Hartmann17}. In these methods, the environmental effect is represented by discretizing/approximating the spectral density\cite{Wang01,Wang03}, the influence functional\cite{Makri}, and the correlation function\cite{Tanimura89,Tanimura90,Nakamura,Tang,Hartmann20,Hartmann17} at some stage.  Thus, it would be necessary to increase terms and/or equations obtained by the discretization until the evaluation converges. However, the number of terms and/or equations that are needed to reach convergence strongly depends on the environmental temperature, cut-off frequency, and time region to be simulated\cite{Hartmann19}.  Obtaining convergence over a wide range of time intervals up to the stationary state is challenging for various environmental temperatures and long correlation time; thus, this will be explored in future studies.

\section{\label{sec:level8}Conclusion}
We studied the reduced dynamics of excitation energy transfer through interacting sites that were partially coupled with a bosonic environment. Using the GA beyond the BMA and the SA, we found the dynamics to asymptotically approach the stationary state near the expected Gibbs state with a trace distance of the order of system--environment interaction. Comparisons with and without the SA revealed the role of the SA in reducing the oscillation of the non-Markovian dynamics.  Specifically, the difference persists for the long-time region in a high-temperature environment. 

These findings suggest that it is feasible and plausible to adopt the GA without resorting to the BMA and the SA.  As for the stationary state specifically in the low-temperature region, the evaluation by the GA with the SA is still a reasonable choice.
\begin{acknowledgments}
This work was supported in part by the MEXT KAKENHI Grant-in-Aid for Scientific Research on Innovative Areas Science of hybrid quantum systems Grant 18H04290, JSPS KAKENHI Grant 15K05200, 22K03467, and the researcher exchange promotion program of the Research Organization of Information and Systems.
\end{acknowledgments}
\vskip 1truecm
\appendix
\section {\label{sec:levelA}Derivation of the time convolutionless (time-local) type of master equation}
We provide a derivation of the time convolutionless (time-local) type of master equation in the Schr\"odinger picture, Eq.~(\ref{eqn:21}), which describes the time evolution of the system of interest extracted from that of the total system given by the Liouville--von Neuman equation,
\begin{eqnarray}
\frac{d}{dt} W(t)=-i \el W(t), \label{eqn:M1}
\end{eqnarray}
where \(W(t)\) denotes the density operator for the total system and \(\el\) the Liouville operator representing the operation \(\el A=\frac{1}{\hbar} [\ch, A] \) for the Hamiltonian of the total system $\ch$ when acting on an arbitrary operator \(A\). We consider dividing the Hamiltonian into unperturbed and perturbed parts denoted $\ch_{0}$ and $\ch_{1}$, respectively. We denote the corresponding Liouville operator as \(\el_{\mu} A=\frac{1}{\hbar} [\ch_{\mu}, A]\) with $\mu=\{0,1\}$.

To extract the time evolution of the system of interest from Eq.~(\ref{eqn:M1}), we use the projection operator method\cite{Kubo,Hanggi,HSS,KTH,STH,CS,FA,US,Breuer}, which is summarized as follows: We introduce a projection operator, \(\cp\), which satisfies the idempotent property \(\cp^2=\cp\) to divide Eq.~(\ref{eqn:M1}) into a relevant part $\cp W(t)$ and an irrelevant part $\cq W(t) (\equiv(1-\cp) W(t))$. When we reduce them to a single equation $\cp W(t)$ and define the projection operator as $\cp W(t)=\rho_{{\rm E}} {\rm Tr}_{\rm E}W(t)$, we obtain the master equation for the reduced density operator, namely $\rho(t)={\rm Tr}_{\rm E}W(t)$, where $\rho_{\rm E}$ denotes the environment density operator and ${\rm Tr}_{\rm E}$ the partial trace operation over environment variables. We present a detailed derivation below.

For the formal solution of Eq.~(\ref{eqn:M1}), given as \(W(t)=U(t) W(0)\) with \(U(t)=\exp[-i \el t]\), we separate the relevant part \(\cp\) and the irrelevant part \(\cq (\equiv 1-\cp)\) of $U(t)$, 
\begin{equation}
x(t) \equiv \cp U(t), \;\;\; y(t)\equiv \cq U(t), \label{eqn:M2}
\end{equation} 
and derive their simultaneous differential equations
\begin{eqnarray}
\frac{d}{dt} x(t)&=&\cp (- i \el ) x(t) + \cp (- i \el ) y(t) \; , \label{eqn:M3}\\
\frac{d}{dt} y(t)&=&\cq (- i \el ) x(t) + \cq (- i \el ) y(t) \; .\label{eqn:M4}
\end{eqnarray}
We obtain the formal solution of the irrelevant part \(\cq\) as
\begin{eqnarray}
y(t)&=&\int_{0}^{t} e^{-\cq i \el (t-\tau)} \cq (-i \el) x(\tau) d\tau + e^{-\cq i \el t} \cq.
\label{eqn:M5}
\end{eqnarray}
The rewriting of \(x(\tau)\) in Eq.~(\ref{eqn:M5}) with \(x(t)\) and \(y(t)\) using the relation 
\begin{equation}
x(\tau)=\cp e^{ i \el (t-\tau)} e^{ -i \el t}=\cp e^{ i \el (t-\tau)} (x(t)+y(t)),
\label{eqn:M6}
\end{equation}
leads to the formal solution of \(y(t)\) in the form
\begin{equation}
y(t)=\theta(t)^{-1}( (1-\theta(t)) x(t)+e^{-\cq i \el t} \cq),
\label{eqn:M7}
\end{equation}
where we define
\begin{equation}
\theta(t)=1-\int_{0}^{t} e^{-\cq i \el \tau} \cq (-i \el) \cp e^{i \el \tau} d\tau \equiv 1-\sigma(t).
\label{eqn:M8}
\end{equation}
Substitution of the formal solution of \(y(t)\) into Eq.~(\ref{eqn:M3}) gives
\begin{equation}
\frac{d}{dt} x(t)=\cp (-i \el) \theta(t)^{-1} x(t) +\cp (-i \el) \theta(t)^{-1} e^{\cq (-i \el) t} \cq.
\label{eqn:M9}
\end{equation}
Using the relations \(\theta(t)^{-1}=\sum_{n=0}^{\infty} \sigma(t)^{n}\), \(\cp \el_{0} =\el_{0} \cp\), \(\cp \cq=0\), and
\begin{equation} 
e^{-\cq i \el t} \cq=e^{-\cq i \el_{0} t} T_{+}{\exp[\int_{0}^{t} dt' e^{i \el_{0} t'} \cq (-i \el_{1}) \cq e^{-i \el_{0} t'}]},
\label{eqn:M10} 
\end{equation} 
the first term on the right-hand side of Eq.~(\ref{eqn:M9}) is rewritten as
\begin{eqnarray} 
\cp (-i \el) \theta(t)^{-1} x(t) && \nonumber \\
&&\hspace{-3cm}=\cp (-i \el) x(t)+ \cp (-i \el) \sigma(t) x(t) + \cdots, \nonumber \\
&&\hspace{-3cm} =\cp (-i \el) x(t)+ \sum_{n=2}^{\infty} k_{n}(t) x(t), \label{eqn:M11} 
\end{eqnarray} 
where we define
\begin{eqnarray} 
k_{n}(t) &\equiv& \int_{0}^t dt_1 \int_{0}^{t_1} dt_2 \cdots \int_{0}^{t_n-2}dt_n \nonumber \\
\hspace{-1cm}&& e^{-i \el_{0} t} \langle i \incl (t) i \incl (t_{1}) \cdots i \incl (t_n-1) \rangle_{o.c.}e^{i \el_{0} t} \; ,\nonumber \\
\label{eqn:M12}
\end{eqnarray}
with
\begin{equation} 
\incl (t)=e^{i \el_{0} t } \el_{1} e^{-i \el_{0} t } .
\label{eqn:M13} 
\end{equation} 
In Eq.~(\ref{eqn:M12}), $\langle \cdots \rangle_{o.c.}$ means the ``ordered cumulant"\cite{Kubo,Hanggi,HSS,KTH,STH,CS,FA,US,Breuer}; the second-order term is 
\begin{eqnarray} 
 k_{2}(t)=\cp (-i \el_{1}) \int_{0}^{t} \cq (-i \incl (-\tau)) \cp d\tau .
\label{eqn:M14} 
\end{eqnarray}
Multiplying \(W(0)\) from the right on both sides of Eq.~(\ref{eqn:M9}), we obtain 
\begin{equation}
\frac{d}{dt} \rho(t)={\rm Tr_{E}}[ (-i \el) \rho_{E} \rho(t)] + \Psi(t) + {\cal I}(t) ,
\label{eqn:M15}
\end{equation}
with
\begin{eqnarray}
 \Psi(t) &\equiv& \sum_{n=2}^{\infty} \psi_{n}(t) \rho(t), \label{eqn:M16}\\
{\cal I}(t) &\equiv& Tr_{\rm E}[ (-i \el) \theta(t)^{-1} e^{\cq (-i \el) t} \cq W(0)]. \label{eqn:M17}
\end{eqnarray}
When we assume the factorized initial condition between the system and the environment satisfies \(\cq W(0)=0\), we find the inhomogeneous term ${\cal I}(t)=0$. Moreover, assuming ${\rm Tr_{E}}(\ch_{1} \rho_{{\rm E}})=0$ where $\ch_{1}$ is denoted as the multiplicative of the system and environment operators, we find the first term in Eq.~(\ref{eqn:M15}) is rewritten as ${\rm Tr_{E}}[ (-i \el)\rho_{E}\rho(t)]={\rm Tr_{E}}[ (-i \el_{0})\rho_{E}\rho(t)]$. Denoting \(\ch_{0}=\ch_{\rm S}+\ch_{\rm E}\) with the system Hamiltonian $\ch_{\rm S}$ and the environmental one $\ch_{\rm E}$, and on the condition of $[\ch_{\rm E}, \rho_{E}]=0$, we obtain Eq.~(\ref{eqn:21}). Using Eqs.~(\ref{eqn:M14}) and (\ref{eqn:M16}), we obtain the second order of the dissipator Eq.~(\ref{eqn:23}).
\section{\label{sec:levelB}Coefficient matrices for non-Markovian dynamics}
For the $2$-site model ($N=2$), the matrices of coefficients in Eqs.~(\ref{eqn:vp}) and (\ref{eqn:vc}) are given by
\begin{eqnarray}
\Gamma_{P,\alpha}(t)&=&\begin{bmatrix}
0 & 0 & 0 \\
0&\eta_{+,\alpha}(t)&\eta_{-,\alpha}(t)\\
0&-\eta_{+,\alpha}(t)&-\eta_{-,\alpha}(t)
\end{bmatrix},\label{eqn:A1}\\
\Gamma_{PC,\alpha}&=&\Gamma_{PC,{\rm sys}}
+\begin{bmatrix}
0&0\\
-\gamma_{1,\alpha}(t)&-\gamma_{1,\alpha}^{*}(t)\\
\gamma_{1,\alpha}(t)&\gamma_{1,\alpha}^{*}(t)
\end{bmatrix},\label{eqn:A2}\\
\Gamma_{C,\alpha}(t)&=&
\Gamma_{C,{\rm sys}}
-\begin{bmatrix}
\gamma_{2,\alpha}(t) & 0 \\
0&\gamma_{2,\alpha}^{*}(t)
\end{bmatrix}, \label{eqn:A3}\\
\Gamma_{CP,\alpha}&=&
\Gamma_{CP,{\rm sys}}
+\begin{bmatrix}
-\gamma_{1,\alpha}^{*}(t)&0&\gamma_{3,\alpha}(t)\\
-\gamma_{1,\alpha}(t)&0&\gamma_{3,\alpha}^{*}(t)
\end{bmatrix},
\label{eqn:A4}
\end{eqnarray}
where $\alpha=G,L$ indicate matrices pertaining to the GA and the LA, respectively, and the first terms in Eqs.~(\ref{eqn:A2})--(\ref{eqn:A4}) denote the matrices of coefficients describing the contributions from the system Hamiltonian; more specifically,
\begin{eqnarray}
\Gamma_{PC,{\rm sys}}&=&i V_{12} \begin{bmatrix}
1& -1 \\
-1&1\\
0&0
\end{bmatrix},\\
\Gamma_{C,{\rm sys}}&=&i (\omega_{1}-\omega_{2})\begin{bmatrix}
-1 & 0 \\
0&1
\end{bmatrix}\;, \\
\Gamma_{CP,{\rm sys}}&=&i V_{12} \begin{bmatrix}
1& -1 &0\\
-1&1&0
\end{bmatrix}.
\end{eqnarray}
In the following, we present explicit expressions for the elements of these matrices.
\subsection{\label{sec:levelB1}Global Approach beyond the secular approximation}
The elements of the matrices for the GA coefficients are given by 
\begin{eqnarray}
\eta_{\pm,G}(t)&\equiv& \mp 2(\sin^2\theta \; {\rm Re}[\Phi(\pm \lambda_{13},t)]
+\cos^2\theta \; {\rm Re}[\Phi(\pm \lambda_{23},t)])\nonumber \\
\gamma_{1,G}(t)&\equiv&\frac{V_{12}}{D_{m}}
(\Phi(\lambda_{13},t)-\Phi(\lambda_{23},t)),\nonumber\\
\gamma_{2,G}(t)&\equiv&\sin^2\theta \;\Phi_{S}(-\lambda_{13},t)+\cos^2\theta\;\Phi_{S}(-\lambda_{23},t)),\nonumber\\
\gamma_{3,G}(t)&\equiv&\frac{V_{12}}{D_{m}}
(\Phi(-\lambda_{13},t)-\Phi(-\lambda_{23},t)), \label{eqn:b8} 
\end{eqnarray}
with $\sin^2\theta=(1-\frac{\omega_{1}-\omega_{2}}{D_{m}})/2$, $\cos^2\theta=(1+\frac{\omega_{1}-\omega_{2}}{D_{m}})/2$, and $\lambda_{nm}\equiv\frac{1}{\hbar}(\lambda_{n}-\lambda_{m})$ representing the frequency difference between the $n$-th and $m$-th eigenvalues of the unperturbed Hamiltonian, and
\begin{eqnarray}
\Phi(\mu,t)\nonumber \\
&&\hspace{-0.7cm}\equiv\int_{0}^{t} d\tau \int_{0}^{\infty} d \nu J(\nu) \{ (1+ 2 n(\nu) )\cos \nu \tau - i \sin \nu \tau \} e^{i \mu \tau} \nonumber \\
&&\hspace{-0.7cm}=\Phi_{S}(-\mu,t)^{*} , \label{eqn:b9} 
\end{eqnarray}
defining $n(\nu)=\frac{1}{e^{\beta \hbar \nu}-1}$ and the spectral density $J(\nu)$ as
\begin{equation}
J(\nu)\equiv\sum_{k}g_{k}^2 \delta(\nu-\omega_{k}).
\label{eqn:b10} 
\end{equation}
When we assume the spectral density to be Ohmic with an exponential cutoff function, $J(\nu) \equiv s\,\nu \,e^{-\nu/\Omega_{c}}$, denoting the strength of system--environment interaction by $s$ and the cut-off frequency by $\Omega_{c}$, we obtain 
\begin{eqnarray}
\Phi(\mu,t) =\int_{0}^{t} d\tau \frac{1}{2} (D_{1}(\tau) - i D_{2}(\tau)) e^{i \mu \tau} , \label{eqn:b11} 
\end{eqnarray}
where $D_{1}(\tau)$ and $D_{2}(\tau)$ are termed the {\it noise} and {\it dissipation} kernels \cite{Clos,Guarnieri16} defined as
\begin{eqnarray}
D_{1}(\tau)
&=&2 \int_{0}^{\infty} d \nu J(\nu) (1+ 2 n(\nu) )\cos \nu \tau \nonumber \\
&=&2 s (-\Omega_{c}^2 \frac{(\Omega_{c} \tau)^2-1}{[1+(\Omega_{c} \tau)^2]^2} \nonumber \\
&&\hspace{0.5cm} + \frac{2}{\beta^2} \Re[\psi' \Big(\frac{\beta \Omega_{c}
+i \Omega_{c} \tau +1}{\beta \Omega_{c}}\Big)]), \\
D_{2}(\tau)&=& \frac{ 4 s \tau \Omega_{c}^3}{[1+(\Omega_{c} \tau)^2]^2},
\end{eqnarray}
with $\psi'(z)$ denoting the derivative of the Euler digamma function $\psi(z)=\Gamma'(z)/\Gamma(z)$.
\subsection{\label{sec:levelB2}Local Approach beyond the secular approximation}
Omitting the interaction between sites in the dissipator, we obtain the master equation for the LA beyond the SA. In this instance, the eigenvalues $\{\lambda_{1},\lambda_{2},\lambda_{3}\}$ reduce to the energy of each site, specifically, $\hbar \{\omega_{1},\omega_{2},\omega_{3}\}$, and the corresponding eigenstates $\{|e_{n}\rangle\}$ for $n=1,2,3$ are unit vectors of the site basis. The time-dependent coefficients $\Gamma_{\mu,L}(t)$ for $\{\mu\}=\{P,PC,CP,C\}$ are obtained with the elements given as
\begin{eqnarray}
\eta_{\pm,L}(t)&\equiv& \mp 2{\rm Re}[\Phi(\pm \lambda_{23},t)], \gamma_{1,L}(t) \equiv 0,\nonumber \\
\gamma_{2,L}(t)&\equiv&\Phi_{S}(-\lambda_{23},t),\,\gamma_{3,L}(t)\equiv 0, \label{eqn:bsl} 
\end{eqnarray}
Comparing Eqs.~(\ref{eqn:b8}) and (\ref{eqn:bsl}), we find that the dissipator of the LA reasonably corresponds to that of the GA for $V_{12}=0$ because at this value $\sin^2\theta=0$, and $\cos^2\theta=1$. We also find that the dissipator by the LA reduces to the GKSL form with time-dependent coefficients.
\section{\label{sec:levelC}Coefficient matrices under BMA}
We obtain the coefficients in the BMA by setting the upper limit of Eq.~(\ref{eqn:b9}) to infinity and introducing $\Gamma_{\mu, \alpha} \equiv \Gamma_{\mu, \alpha}(\infty)$ for $\{\mu\}=\{P,PC,CP,C\}$ and $\{\alpha\}=\{G,L\}$, corresponding to the long-time limit, which requires obtaining $\Phi(\mu,\infty)$. For the Ohmic spectral density, its analytic form is given by
\begin{eqnarray}
\Phi(\mu,\infty) &=& \int_{0}^{\infty} d\tau \frac{1}{2}
(D_{1}(\tau) - i D_{2}(\tau)) e^{i \mu \tau} , \nonumber \\
&&\hspace{-1.5cm}=\lim_{\varepsilon \to 0} \int_{0}^{\infty} d\tau
\int_{0}^{\infty} d \nu J(\nu) \{n(\nu) e^{i (\nu+\mu) \tau} \nonumber \\
&&\hspace{2cm}+(1+ n(\nu)) e^{-i(\nu-\mu) \tau} \} e^{-\varepsilon t}\nonumber \\
&&\hspace{-1.5cm} =\int_{0}^{\infty} d \nu J(\nu) \{ n(\nu) F_{+}(\mu, \nu)
+ (1+n(\nu))F_{-}(\mu, \nu) \}\nonumber \\
\label{eqn:C1}
\end{eqnarray}
with
\begin{eqnarray}
F_{\pm}(\mu,\nu) \equiv \pm i {\cal P} \frac{1}{\nu\pm\mu}+\pi \delta(\nu\pm\mu),
\end{eqnarray}
which is obtained using the relation
\begin{eqnarray}
\lim_{\varepsilon \to 0} \frac{1}{t \mp i \varepsilon}
={\cal P}\frac{1}{t} \pm i \pi \delta(t),
\end{eqnarray}
where ${\cal P}$ means Cauchy's principal value. 
\section{\label{sec:levelD}Numerical evaluations for the Local Approach}
\begin{figure}[ht]
\includegraphics[scale=0.25]{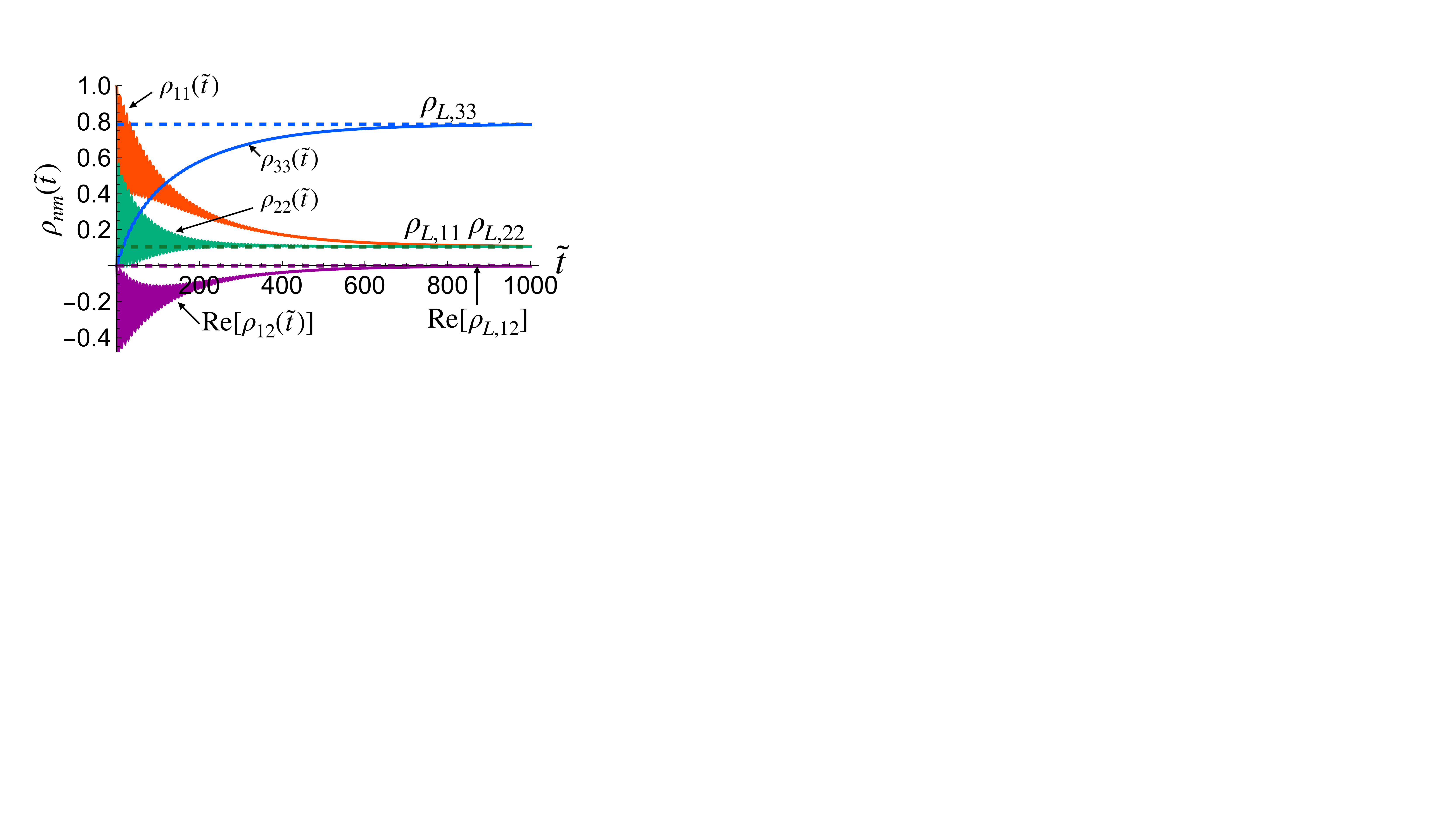}
\caption{Time evolution of each element of \(\rho({\tilde t(=\omega_{2} t} ))\) obtained by the LA beyond the BMA and the SA with the same settings as given in Fig.~\ref{fig:fig1}. We find that the stationary values of the populations at the sites $1$ and $2$ coincide, $\rho_{L,11}=\rho_{L, 22}$, and the stationary value of coherence ${\rm Re}[\rho_{L, 12}]$ vanishes.}
\label{fig:figD1}
\end{figure}

\begin{figure}[ht]
\includegraphics[scale=0.2]{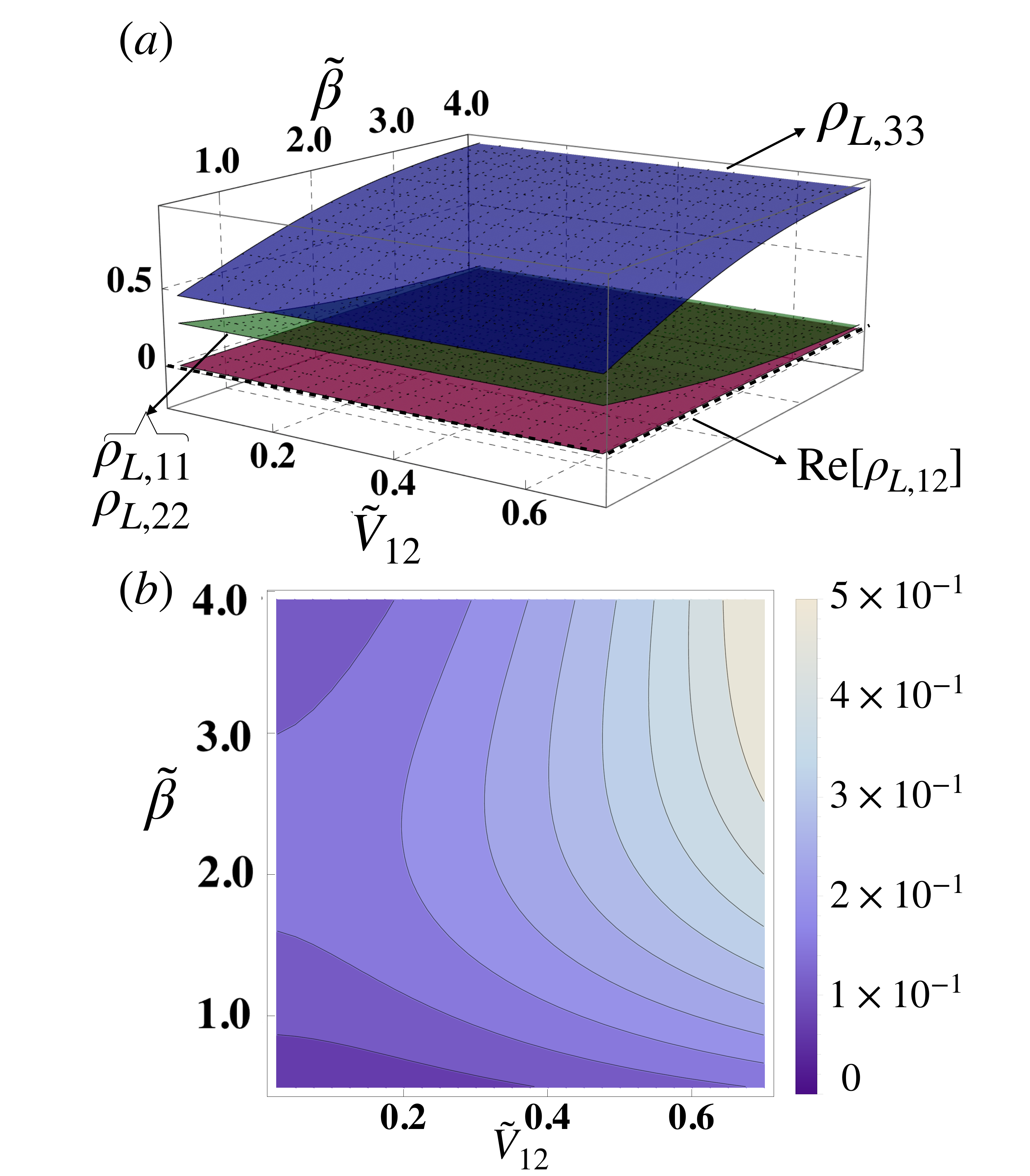}
\caption{(a) Dependence of the stationary values obtained by the LA on the inverse temperature ${\tilde \beta}$ and interaction strength between sites ${\tilde V}_{12}$. (b) Dependence of the trace distance, $D(\rho_{L}, \rho_{{\rm Gibbs}})$, on ${\tilde \beta}$ and ${\tilde V}_{12}$. The difference between the obtained stationary state, $\rho_{L}$, and the Gibbs state for the total relevant system, $\rho_{{\rm Gibbs}}$, is much larger compared with $D(\rho_{G}, \rho_{{\rm Gibbs}})$ [Fig.~\ref{fig:fig2}(b)].}
\label{fig:figD2}
\end{figure}

We display the non-Markovian reduced dynamics obtained by the LA beyond the SA graphically (Fig.~\ref{fig:figD1}). The initial condition and parameters are the same as set for the GA (Fig.~\ref{fig:fig1}). 
We find that the population of sites $1$ and $2$ approach the same respective values $\rho_{L,11}$ and $\rho_{L, 22}$, which is consistent with the coherence ${\rm Re}[\rho_{L, 12}]$ vanishing and therefore the vanishing of the transition between sites $1$ and $2$. These features are supported by an analytic solution of the stationary state obtained using Eq.~(\ref{eqn:C1}) in Eq.~(\ref{eqn:bsl}),
\begin{eqnarray}
\rho_{L,11}&=&\rho_{L,22}=\frac{e^{-\beta \hbar \omega_{2}}}{Z_{L}}, \;\; \rho_{L,33}=\frac{e^{-\beta \hbar \omega_{3}}}{Z_{L}}, \nonumber \\
\rho_{L,12}&=&\rho_{L,21}=0, \label{eqn:D1}\\ \nonumber 
\end{eqnarray}
with $Z_{L}=2 e^{-\beta \hbar \omega_{2}}+e^{-\beta \hbar \omega_{3}}$, which means that both sites $1$ and $2$ are populated in local equilibrium with site $3$.
We present the dependence of the stationary values for the LA on ${\tilde \beta}$ and ${\tilde V}_{12}$ in Fig.~\ref{fig:figD2}(a). From a comparison with Fig.~\ref{fig:fig2}(a), we find the LA does not provide the stationary state of the Gibbs state for the total Hamiltonian ${\ch}_{S}$, which is consistent with conventional studies\cite{Walls,Schwendimann,Carmichael,Gardiner,Cresser}. Evaluating the trace distance between the stationary state $\rho_{L}$ and the Gibbs state for ${\ch}_{S}$, $D(\rho_{L}, \rho_{{\rm Gibbs}})$, its dependence on ${\tilde \beta}$ and ${\tilde V}_{12}$ [Fig.~\ref{fig:figD2}(b)] shows that the values are much larger than those for the GA [Fig.~\ref{fig:fig2}(b)].
\section{\label{sec:levelE}On the positivity of the stationary state}
\begin{figure}[ht]
\includegraphics[scale=0.25]{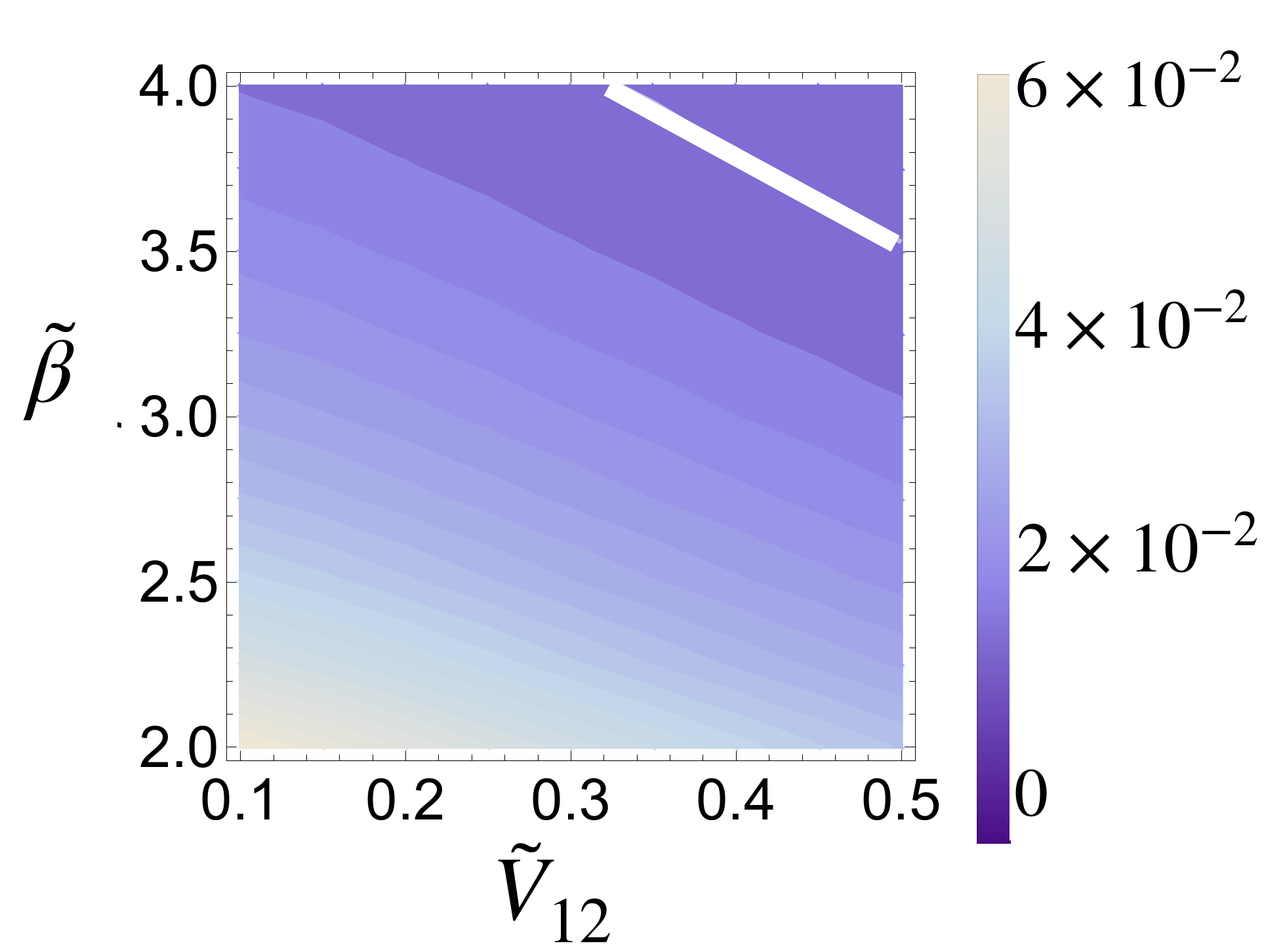}
\caption{Dependence of the smallest eigenvalue of the stationary state obtained by the GA beyond the SA on the inverse temperature ${\tilde \beta}$ and the interaction strength between sites ${\tilde V}_{12}$ with setting ${\tilde \omega}_{1}=5/4$. The white line marks the boundary between positive and negative values; below the line, the eigenvalues are positive and in crossing the boundary change to very small negative values. Other parameters are set the same as given in Fig.~\ref{fig:fig1}. }
\label{fig:figE1}
\end{figure}

When adopting the GA beyond the SA, we find no violation of positivity within the parameter range in Figs.~\ref{fig:fig1}--\ref{fig:fig5}. In the evaluations, we found that the violation of positivity in the short-time region is resolved by including the non-Markovian effect. This means the remaining possibility of violation is in the stationary state which requires us to study its positivity by enlarging the parameter region of the inverse temperature of the environment, ${\tilde \beta}$, the strength of inter-site interaction, ${\tilde V}_{12}$, and the Bohr frequency of the site $1$, ${\tilde \omega_{1}}$. Changing these parameters, we identify the region for which the smallest eigenvalue of the stationary state is positive (Fig.~\ref{fig:figE1}). We then find violations of positivity occurring at low environment temperatures and large ${\tilde \omega_{1}}$ and large ${\tilde V}_{12}$, although their explicit values are small. Incidentally, the scaled temperature, ${\tilde \beta=4}$, corresponds to about $13$ mK when the scaling Bohr frequency ${\omega_2}$ is assumed to be $7$ GHz, which is much lower than the case in the heat rectification experiment\cite{Senior}.
\\

\end{document}